\documentstyle[amsfonts,aps]{revtex}

\begin{document}



\title{TFD Approach to Bosonic Strings and $D_{p}$-branes
\footnote{Proceedings Contribution for the Second Londrina Winter School 
- Mathematical Methods in Physics, August 25-30, 2002, Londrina - Pr, Brazil.}
}

\author{\footnotesize M. C. B Abdalla$^{*}$ and
A. L. Gadelha}

\address{Instituto de F\'{\i}sica Te\'{o}rica, Unesp, Pamplona 145, S\~{a}o Paulo,
SP, 01405-900, Brazil, $^{*}$mabdalla@ift.unesp.br }

\author{\footnotesize I. V. Vancea}

\address{Departamento de F\'{\i}sica Matem\'{a}tica, Faculdade de Filosofia
Ci\^{e}ncias e Letras de Ribeir\~{a}o Preto, USP, Av. Bandeirantes, 3900,
Ribeir\~{a}o Preto, SP, 14040-901, Brazil}

\maketitle


\begin{abstract}
In this work we explain the construction of the thermal vacuum for the bosonic
string, as well that of the thermal boundary state interpreted as a
$D_{p}$-brane at finite temperature. In both case we calculate the respective
entropy using the entropy operator of the Thermo Field Dynamics theory. We show that
the contribution of the thermal string entropy is explicitly present in the
$D_{p}$-brane entropy. Furthermore, we show that the Thermo Field approach
is suitable to introduce temperature in boundary states.

\end{abstract}

\section{Introduction}
$D_{p}$-branes at finite temperature have been object of intense study. In particular,
they appear as solitonic solutions of Supergravity in the non perturbative limit.
In this limit, the search for an understanding of
its statistical properties has attracted many researchers {\cite{mvm}}.
A plethora of results have been obtained contributing to clarify many
difficulties as, for example, the thermodynamics of black holes and string
gases {\cite{as}}.

However, in the perturbative limit of string theory, the
understanding of the microscopic properties of $D_{p}$-branes at finite 
temperature still waits for a more satisfactory development. With the aim of
changing this picture, a new approach using the framework of Thermo Field
Dynamics (TDF){\cite{tu}} was proposed {\cite{ivv}}-{\cite{su11}} recently. 
This formulation consists in identifying the statistical average of some
operator with a vacuum expectation value in quantum field theory which
must take into account a temperature-dependent state. Indeed, the
basic statement of TFD is, considering an operator $ {\mathcal O} $, 
\begin{equation}
\left\langle {\mathcal O}\right\rangle =\frac{1}{Z\left( \beta \right) }
Tr\left[ \rho {\mathcal O}\right] \equiv \left\langle 0\left( \beta \right)
\left| {\mathcal O}\right| 0\left(\beta \right) \right\rangle. \label {BS} 
\end{equation}
Here, $Z\left( \beta \right) =Tr\left[ \rho \right]$ , 
is the partition function, $\rho$ is the density operator and $\beta=1/k_{B}T$,
for $k_{B}$ being the Boltzmann constant and $T$ the temperature of the system. 
In eq. (\ref{BS}) the temperature dependent state is denoted by
$ |0\left(\beta \right)\rangle $.
The identification (\ref{BS}) is satisfied if we double the degrees of
freedom of the system. To achieve this we need to work in an extended Hilbert
space which consists of a product of two subspaces: the original system,
${\cal H}$, and another one, an auxiliary space denoted by a tilde, 
$\widetilde{{\cal H}}$, independent and with the same characteristics of the
original one. Being $\widehat{{\cal H}}$ the extended space, we have
$\widehat{{\cal H}}={\cal H}\otimes \widetilde{{\cal H}}$. 
By the same characteristics we mean that, for any state or operator associated
to the original Hilbert space we have a partner in the tilde Hilbert
space. So, for all orthonormalized states
$\left| n\right\rangle \in {\cal H}$ one has the states 
$\left|\widetilde{n} \right\rangle \in \widetilde{{\cal H}}$.
In this sense, if $H$ is the hamiltonian operator of the original system and
$\widetilde{H}$ is its equivalent for the auxiliary tilde space, we may
write
\begin{equation}
H\left. \left| n\right\rangle \!\right\rangle =E_{n}\left. \left|
n\right\rangle \!\right\rangle ,\qquad \widetilde{H}\left. \left|
n\right\rangle \!\right\rangle =E_{n}\left. \left| n\right\rangle
\!\right\rangle. 
\end{equation}
Here, we have introduced the notation $\left. \left| n\right\rangle
\!\right\rangle =\left| n,\widetilde{n}
\right\rangle =\left| n\right\rangle \otimes \left| \widetilde{n}
\right\rangle \in \widehat{{\cal H}}$.
Furthermore, all operators ${\cal O}$ and $\widetilde{{\cal O}}$ that act in
the original and auxiliary Hilbert spaces commute.
Under the above considerations, the solution for the identity (\ref{BS}) is
\begin{equation}
\left| 0\left( \beta \right) \right\rangle =\frac{1}{\left( Z\left( \beta
\right) \right) ^{1/2}}\sum_{n}e^{-\frac{\beta }{2}E_{n}}\left| n,\tilde{n}
\right\rangle. 
\end{equation}

An interesting point of
TFD{\footnote{The concepts of TFD, as well as its applications,
are explained in ref. 8 and references therein.}} is that the
temperature-dependent state, can be obtained using a linear canonical
transformation, i.e. a Bogoliubov transformation. To show this we will apply
the formalism to bosonic strings {\cite{gsw}}.
In Sec. 2 we present the basic concepts of the TFD approach. 
In Sec. 3, we construct the boundary states at finite temperature which 
are interpreted as thermal $D_{p}$-brane. In Sec. 4 we discuss the
results.  

\section{Bosonic Strings at Finite Temperature}

Consider the bosonic string in the conformal gauge. The space-time has $D=26$,
the metric is lorentzian and the string spans a world-sheet
that can be described by two parameters. Let $X^{\mu}$ be the string
position vector,  from the view point of the world-sheet. It can be
thought as a 2-dimensional  bosonic free field
$X^{\mu}\left(\tau,\sigma \right)$, with $\tau$ and $\sigma$ the parameters of
the world-sheet. This field satisfies a massless 2-dimensional
Klein-Gordon equation.

We can impose periodic conditions to the solution for this equation
$X^{\mu }\left( \tau ,0\right)=X^{\mu }\left( \tau ,\pi \right)$ defining, in 
this way, the closed bosonic string.
The solution that satisfies these conditions has a Fourier expansion in
terms of right- and left-modes with coefficients $\alpha_{k}^{\mu}$ and
$\beta_{k}^{\mu}$, respectively.

The vanishing of the energy-momentum tensor characterizes the constrained
system. To avoid spurious states in the vector space of the quantized
system, we choose the light-cone gauge,
$X^{\stackrel{+}{-}}=X^{0}\underline{+}X^{25}$,
with the transversal modes described by $X^{i}$, $ i=1,2,...24$. In this way,
we have the following commutation relations
\begin{equation}
\left[ \alpha _{m}^{i},\alpha _{n}^{j}\right] = \left[ \beta _{m}^{i},\beta
_{n}^{j}\right] =m\delta ^{ij}\delta _{m+n,0}, \qquad
\left[ \alpha _{m}^{i},\beta _{n}^{j}\right]=0.
\end{equation} 
These oscillators-like operators can be redefined as
\begin{equation}
A_{k}^{i}=\frac{1}{\sqrt{k}}\alpha _{k}^{i},\qquad A_{k}^{i\dagger }=
\frac{1}{\sqrt{k}}\alpha _{-k}^{i},
\qquad B_{k}^{i} =\frac{1}{\sqrt{k}}\beta _{k}^{i},\qquad B_{k}^{i\dagger }=
\frac{1}{\sqrt{k}}\beta _{-k}^{i},
\end{equation}
for $k>0$, in such a way that we have the vacuum defined by
\begin{equation}
A_{k}^{i}\left| 0\right\rangle \!=B_{k}^{i}\left| 0\right\rangle \!=0, 
\end{equation}
where $ \left| 0\right\rangle \!=\left| 0\right\rangle \!_{\alpha }\left|
0\right\rangle \!_{\beta } $, and the subscripts $\alpha$ and $\beta$
refers to the left- and right-modes for the closed string.

To apply the TFD we need firstly to double   the space. So we have
an auxiliary system called ``tilde system'' which means that beside the
hamiltonian of the original system, $H$, we have also the tilde hamiltonian:
\begin{equation}
H = \sum_{n>0}^{\infty }\left( nA_{n}^{\dagger }\cdot
A_{n}+nB_{n}^{\dagger }\cdot B_{n}\right),
\qquad
\widetilde{H}=\sum_{n>0}^{\infty }\left( n\tilde{A}_{n}^{\dagger }\cdot 
\tilde{A}_{n}+n\tilde{B}_{n}^{\dagger }\cdot \tilde{B}_{n}\right) .
\end{equation}
The oscillators-like operators for the doubled system satisfy the following
algebra
\begin{equation}
\left[ A_{k}^{i},A_{m}^{j\dagger }\right] =
\left[ \tilde{A}_{k}^{i},\tilde{A}_{m}^{j\dagger }\right] =
\left[ B_{k}^{i},B_{m}^{j\dagger }\right]=
\left[ \tilde{B}_{k}^{i},\tilde{B}_{m}^{j\dagger }\right]=
\delta _{km}\delta ^{ij},
\end{equation}
and zero for the rest.
The vacuum of the total system is defined by
\begin{equation}
A_{k}^{i}\left. \left| 0\right\rangle \!\right\rangle =B_{m}^{j}\left.
\left| 0\right\rangle \!\right\rangle =\tilde{A}_{k}^{i}\left. \left|
0\right\rangle \!\right\rangle =\tilde{B}_{m}^{j}\left. \left|
0\right\rangle \!\right\rangle =0.
\end{equation}
In the above expression we have used the following notation 
$\left. \left| 0\right\rangle \!\right\rangle =\left. \left| 0\right\rangle
\!\right\rangle _{\alpha }\left. \left| 0\right\rangle \!\right\rangle
_{\beta }$.

As was have already mentioned, the temperature-dependent state can be obtained
as a Bogoliubov transformation
\begin{equation}
\left| 0\left( \beta \right) \right\rangle =e^{-iG}\left. \left|
0\right\rangle \!\right\rangle. \label{vbt}
\end{equation} 
In the case of closed bosonic string, we have two independent original
subsystems, for the left-  and right-modes. So we need a transformation
generator that acts on both modes. The expression for that generator is
given by 
\begin{equation}
G=G^{\dagger}=\sum_{k}\left( G_{k}^{\alpha }+G_{k}^{\beta }\right) , 
\label{gera}
\end{equation} 
with
\begin{equation}
G_{k}^{\alpha } =
-i\theta_{k} \left( A_{k}\cdot \tilde{A}_{k}-
\tilde{A}_{k}^{\dagger }\cdot A_{k}^{\dagger }\right) ,
\qquad
G_{k}^{\beta }=
-i\theta_{k} \left( B_{k}\cdot \tilde{B}_{k}-\tilde{B}
_{k}^{\dagger }\cdot B_{k}^{\dagger }\right).
\label{gera2}
\end{equation}
Once more, the superscripts denote left- and right-modes. The dot means the
euclidian scalar product in the transverse space-time. With that generator
we obtain explicitly the temperature dependent state as
\begin{eqnarray}
\left| 0\left( \beta \right) \right\rangle &=&
\prod_{k}\left[ \cosh\left(\theta_{k} \right) \right]^{-2Tr\delta ^{ij}}
e^{\tanh \left( \theta_{k} \right)
\left( \widetilde{A}_{k}^{\dagger }\cdot A_{k}^{\dagger }\right) }
e^{\tanh \left( \theta_{k} \right) 
\left( \tilde{B}_{k}^{\dagger }\cdot B_{k}^{\dagger }\right) }
\left. \left|0\right\rangle \!\right\rangle
\\
&=& \prod_{k}\left[ u_{k}\right]^{-2tr\delta ^{ij}}
e^{\frac{v_{k}}{u_{k}}
\left( \widetilde{A}_{k}^{\dagger }\cdot A_{k}^{\dagger }\right) }
e^{\frac{v_{k}}{u_{k}} 
\left( \tilde{B}_{k}^{\dagger }\cdot B_{k}^{\dagger }\right) }
\left. \left|0\right\rangle \!\right\rangle . 
\label{vaexp}
\end{eqnarray}
with
$u_{k} =u\left(\theta_{k}\right)=\cosh \left( \theta _{k}\right)$ 
and 
$v_{k} =v\left(\theta_{k}\right)=\mbox{senh} \left( \theta _{k}\right)$.
The transformations for the $A$ operators are given as follows:
\begin{eqnarray}
A_{k}^{i}\left( \theta \right) &=&
e^{-iG}A_{k}^{i}e^{iG} =
u_{k}A_{k}^{i}-v_{k} \widetilde{A}_{k}^{i\dagger },
\label{ot1}
\\
\widetilde{A}_{k}^{i}\left( \theta\right)&=&
e^{-iG}\widetilde{A}_{k}^{i}e^{iG}=
u_{k}\widetilde{A}_{k}-v_{k}A_{k}^{\dagger },
\label{ot2}
\end{eqnarray}
and similarly for the $B$ operators.

Observing the structure of thermal-dependent state (\ref{vbt})-(\ref{vaexp})
and the expressions for the transformed operators (\ref{ot1})-(\ref{ot2}),
we can introduce the so called tilde conjugation rules which maps the
operators living in the original subspace in to those of the auxiliary
(tilde) space and {\it vice-versa}
\begin{eqnarray}
\left( AB\right) \widetilde{^{{}}}&=&\widetilde{A}\widetilde{B},\\
\left( c_{1}A+c_{2}B\right) \widetilde{^{{}}}&=&\left( c_{1}^{*}\tilde{A}
+c_{2}^{*}\tilde{B}\right) ,\\
\left( A^{\dagger }\right) \widetilde{^{{}}}=\tilde{A}^{\dagger }&,&
\qquad 
\left( \tilde{A}\right) \widetilde{^{{}}}=A,\\
\left( \left| 0\left( \theta \right) \right\rangle \right) \widetilde{^{{}}}
=\left| 0\left( \theta \right) \right\rangle &,&
\qquad
\left( \left\langle 0\left( \theta \right) \right| \right) \widetilde{^{{}}}
=\left\langle 0\left( \theta \right) \right| ,
\end{eqnarray}
with $A$ and $B$ bosonic operators and $c_{1}$, $c_{2}\in {\Bbb C}$.

Now it is possible to define the hamiltonian of the total system. First we
note that as the auxiliary system independs of the original one,
the introduction of the former does not change the dynamics of the latter,
in such a way that the total hamiltonian, denoted by $\widehat{H}$, does
not contain terms like $A\widetilde{A}$. Furthermore, as the tilde system
is identical to the original one, its dynamics has to be the same. These
conditions means that the Heisenberg equations for the free bosonic fields
and its tilde counterpart must be the same. Joining these statements together
with the tilde conjugation rules we find the total hamiltonian, $\widehat{H}$
\begin{equation}
\widehat{H}=H-\widetilde{H}. 
\end{equation}
The hamiltonian operator that satisfies the condition for the extended system
is given by 
\begin{equation}
\hat{H} = H-\tilde{H}=
\sum_{n>0}^{\infty }\left( nA_{n}^{\dagger }\cdot
A_{n}+nB_{n}^{\dagger }\cdot B_{n}\right) -\sum_{n>0}^{\infty
}\left( n\tilde{A}_{n}^{\dagger }\cdot \tilde{A}_{n}+n
\tilde{B}_{n}^{\dagger }\cdot \tilde{B}_{n}\right).
\end{equation}
When the approach is applied to equilibrium systems the transformation generator
commutes with the hamiltonian of the total system.
A direct implication of this statement is that the total hamiltonian is
invariant under transformation generated by (\ref{gera}).

The Bogoliubov transformation is canonical, in such a way that the commutation
relations, for the oscillators-like operators, remain unchanged.
The vacuum of the transformed system is just the temperature-dependent state and
for this reason it is called thermal vacuum.

\section{ Bosonic $D_{p}$-branes at Finite Temperature}

$D_{p}-$branes are extended objects that can have open strings endpoints
attached to them{\footnote{For the boundary states construction of
$D_{p}$-brane we follow Ref. 10.}}. 
This can be realized if we impose Neumann boundary conditions in the
parallel directions and Dirichlet boundary conditions in the transversal
directions. For an open string with one end ($\sigma=0$ for instance)
attached to the object, we have
\begin{eqnarray}
\left. \partial _{\sigma }X^{a}\right| _{\sigma =0} &=&0,\qquad a=1,...,p,
\\
\left. X^{i}\right| _{\sigma =0} &=&x^{i},\qquad i=p+1,...24,
\end{eqnarray}
where $a=1,...,p$ are the parallel directions to the $D_{p}$-brane and
$i=p+1,...24$ are the transversal ones.
This description is in the open string channel.
A conformal transformation can be used to map these conditions, imposed
for the open string solution, to the following conditions imposed to the
closed string
\begin{eqnarray}
\left. \partial _{\tau }X^{a}\right| _{\tau =0} &=&0,\qquad a=1,...p, \\
\left. X^{i}\right| _{\tau =0} &=&y^{i},\qquad i=p+1,...,24.
\end{eqnarray}
After the quantization of the system the above conditions for closed string
can be read, in terms of the left- and right-modes operators, as the
follow operator equations
\begin{equation}
\left( A_{k}^{\mu}+S^{\mu}\,\!\!_{\nu}B_{k}^{\nu\dagger }\right)
\left| B_{X}\right\rangle =0, \qquad
\left( A_{k}^{\mu\dagger } + S^{\mu}\,\!\!_{\nu}B_{k}^{\nu}\right) 
\left| B_{X}\right\rangle  = 0,
\label{ec1}
\end{equation} 
\begin{equation}
\widehat{p}^{a}\left| B_{X}\right\rangle \!=
\left( \widehat{q}^{i}-y^{i}\right) \left| B_{X}\right\rangle \!=0,
\label{ec2}
\end{equation}
for $k>0$, and $\widehat{p}^{a}$ and $\widehat{q}^{i}$ are the momentum
and coordinate operators for the center-of-mass, respectively. 
$S^{\mu \nu}\equiv\left(\delta^{ab}, -\delta^{ij}\right)$
was defined in order to obtain a compact expression.
Expressions (\ref{ec1})-(\ref{ec2}) define what we call boundary
states denoted by $\left| B_{X}\right\rangle $. The state that satisfies
these expressions can be written as 
\begin{equation}
\left| B_{X}\right\rangle \!=N_{p}\delta ^{\left( d_{\bot }\right) }
\left(\widehat{q}-y\right) \exp \left[ -\sum_{n=1}^{\infty }
A_{n}^{\dagger }\cdot S\cdot B_{n}^{\dagger }\right] \left| 0\right\rangle\!,
\label{db}
\end{equation}
where $N_{p}$ is a normalization constant, the delta function localizes
the brane in the transverse space and the vacuum is the closed
bosonic string vacuum.

Following the TFD approach, the doubling of the space implies that
beside the expressions (\ref{ec1})-(\ref{ec2}) we have another set
obtained by the tilde conjugations rules. The boundary state
(\ref{db}) has a partner in the tilde space in such a way that the
state considered now in the extended space in given by
\begin{equation}
\left. \left| B_{X}\right\rangle \!\right\rangle =\left| B_{X}\right\rangle 
\widetilde{\left| B_{X}\right\rangle }.
\end{equation}
The temperature-dependent boundary state, interpreted here as a thermal
$D_{p}$-brane can be obtained as before, performing the Bogoliubov
transformation as 
\begin{equation}
\left| B_{X}\left( \theta \right) \right\rangle =e^{-iG}\left. \left|
B_{X}\right\rangle \!\right\rangle, 
\end{equation}
where the generator in given by expressions (\ref{gera})-(\ref{gera2}).
The transformation leads us to an explicit expression for the thermal
$D_{p}$-brane. Namely,
\begin{eqnarray}
\left| B_{X}\left( \theta \right) \right\rangle \! &=&N_{p}^{2}\delta
^{\left( d_{\bot }\right) }\left( \widehat{q}-y\right) \delta ^{\left(
d_{\bot }\right) }\left( \widehat{\widetilde{q}}-\widetilde{y}\right)  
\nonumber
\\
&&\times \exp \! \left\{ -\sum_{k=1}^{\infty }
\left[A_{k}^{\dagger }\left(\theta\right) \cdot S\cdot B_{k}^{\dagger }
\left( \theta \right)+\widetilde{A}_{k}^{\dagger }\left( \theta \right)
\cdot S\cdot \widetilde{B}_{k}^{\dagger }\left( \theta \right) 
\right]\right\}  
\left| 0\left( \theta \right) \right\rangle \!.
\label{tdb}
\end{eqnarray}
Here, we assume that the normalization constant, as well as the
momentum and coordinate operators, do not change by thermal effects. 
The boundary conditions that define the boundary state are now given by
\begin{equation}
\left[ A_{k}^{\mu}\left( \theta \right) +
S^{\mu}\,\!\!_{\nu}B_{k}^{\nu\dagger }\left( \theta \right)
\right] \left| B_{X}\left( \theta \right) \right\rangle =0,
\qquad
\left[ A_{k}^{\mu\dagger}\left( \theta \right) +
S^{\mu}\,\!\!_{\nu}B_{k}^{\nu }\left( \theta \right)
\right] \left| B_{X}\left( \theta \right) \right\rangle =0.
\end{equation} 
and a similar set is obtained by the tilde conjugation. Note that the thermal
boundary state is invariant by the tilde conjugation rules.

\section{ TFD Entropy Operator}

In his original paper, Takahashi and Umezawa defined an operator that
when its expectation value is calculated in the thermal vacuum multiplied
by the Boltzmann constant, in the Stirling approximation, results the general
formula for the entropy. For this reason this operator was called the
entropy operator. 
 
In our case the expression for the entropy operator is given by
\begin{equation}
K=K^{\alpha} + K^{\beta}=
\sum_{k}\left(K_{k}^{\alpha} + K_{k}^{\beta}\right).
\label{kst}
\end{equation}
The upper indices refer to the right- and left-modes as before.
Explicitly we have
\begin{equation}
K^{\alpha}=-\sum_{k}\left[ A_{k}^{\dagger }\cdot A_{k}
\log \left(g \sinh^{2}\left( \theta_{k}\right) \right) -
A_{k}\cdot A_{k}^{\dagger }\log \left(1+g\sinh^{2}\left(
\theta_{k} \right) \right) \right],
\end{equation}
and the same for $K^{\beta}$ by the change of $A$ by $B$ operators. Here
$g=tr\delta^{ij}$. 
As in the original case, the result is given in terms of the thermal vacuum
expectation value of the number operators for the left- and right-modes.
Both modes contribute by the same quantity, 
\begin{equation}
n_{k}=\left\langle 0\left( \theta \right) \right| N_{k}^{\alpha}
\left| 0 \left( \theta\right) \right\rangle=
\left\langle 0\left( \theta \right) \right| N_{k}^{\beta}
\left| 0 \left( \theta\right) \right\rangle=
g\sinh^{2}\left( \theta_{k} \right)=
\frac{e^{-\left( k_{B}T\right) ^{-1}\omega _{k}}}
{1-e^{-\left( k_{B}T\right) ^{-1}\omega _{k}}} .
\label{nst} 
\end{equation}
In the case of bosonic string, the thermal vacuum expectation value 
for the entropy operator (\ref{kst}) is interpreted as the entropy of
the string and we find
\begin{equation}
S_{cs}=k_B\left\langle 0\left( \theta \right) \right| K\left| 0\left(
\theta \right) \right\rangle \!=
2k_{B}\left\{ \sum_{k}\left[ \left( g+n_{k}\right)
\log \left( 1+n_{k}\right) -n_{k}\log \left( n_{k}\right) \right] \right\} ,
\label{ecs}
\end{equation}
where  $n_{k}$ is the same given by (\ref{nst}). The factor two comes from
the equal contribution of the right- left-modes of the bosonic closed
string.

We can perform the calculation, for the expectation value of the entropy
operator in the state that represents the $D_{p}$-brane at finite
temperature given by (\ref{tdb}). In this case we find
\footnote{This expression can be obtained from the entropy of a 
$D_{p}$-brane in an external field presented in Ref. 6, by considering a
vanishing external field and a suitable choice of parameters.}
\begin{eqnarray}
S_{Db}=k_{B}\left\langle B_{X}\left( \theta \right) \right| K
\left| B_{X}\left(\theta \right) \right\rangle = 
-2k_{B} \sum_{k} \left\{ \left ( 1+2n_k \right){\cal F}_k
\log \left( \frac{n_k}{1+n_k} \right)\right\}
+ 
\nonumber
\\
-2k_{B}\sum_{k}\left\{ n_k\log \left( {n_k} \right)
-\left( g+n_k \right) \log(1+n_k) \right\},  
\label{edb}
\end{eqnarray}
where ${\cal F}_{k}$ is given by
\begin{eqnarray}
{\cal F}_{k} ={\cal N}^{2}\widetilde{{\cal N}}^{2}\sum_{t_{1}^{1,1},
\cdots,t_{n}^{24,24}}\sum_{s_{1}^{1,1}\cdots s_{n}^{24,24}}
&&\sum_{\rho}\sum_{\sigma }(S_{24,24})^{^{2t_{1}^{24,24}}}\cdots 
\nonumber
\\
&&\times
(S_{1,1})^{^{2t_{n}^{1,1}}}(S_{24,24})^{^{2s_{1}^{24,24}}}\cdots
(S_{1,1})^{^{2s_{n}^{1,1}}}s_{m}^{\rho ,\sigma },
\end{eqnarray}
where we define
${\cal N}\equiv N_{p}\left(
F\right) \delta ^{\left( d_{\perp }\right) }\left( \hat{q}-y\right) $
and
$\widetilde{{\cal N}}\equiv N_{p}\left( F\right)
\delta ^{\left( d_{\perp }\right) }\left( 
\widetilde{\hat{q}}-\widetilde{y}\right) $.
As before, the global factor two comes from the equal contribution of the
independent modes of the closed string. Note that the contribution of the
second term in the above expression, is exactly the entropy of the closed
bosonic string given by (\ref{ecs}).
\section{Conclusions}
In this work we present the construction of thermal vacuum for bosonic
closed string using the TFD approach.
Boundary states at finite temperature were obtained from the imposition of
boundary conditions at the solution of the bosonic thermal string and
these thermal boundary states were interpreted as a thermal $D_{p}$-brane.  
Using a modified entropy operator, we
obtained the entropy for these states as well for the bosonic closed string.
The use of TFD approach seems to be suitable to the study of
microscopic structure, in the perturbative limit at finite
temperature. The result for the $D_{p}$-brane entropy presents an explicit
contribution of the thermal bosonic string. We note the entropy expressions
for both systems ara similar to the usual entropy for bosonic particles. 
With the study of dimensional compacted space-time, the contributions of
winding-modes will be present. Also, we note that the same analysis that has
been done here in the light-cone gauge can be performed in the covariant gauge
by including the ``thermal ghosts'' {\cite{PLAp}}.
We hope that our construction can throw some
light in the microscopic structure of these kind of systems and that the
statistical properties of $D_{p}$-branes can be better understood and clarified. 
\section*{Acknowledgements}
I. V. V. was partially supported by the FAPESP Grant 02/05327-3.

\end{document}